\documentclass[twocolumn,showpacs,preprintnumbers,amsmath,amssymb,superscriptaddress,10pt,aps,prb]{revtex4}
\usepackage{graphicx}
\usepackage{hyperref}
\usepackage{amssymb}
\usepackage{dcolumn}
\usepackage{float}
\usepackage{bm}
\usepackage{amsmath,amssymb,amsthm} 
\usepackage[utf8]{inputenc}
\usepackage{lmodern}
\usepackage{color}

\usepackage{epsfig,amsopn}
\usepackage{graphicx}
\usepackage{epstopdf}

\usepackage{verbatim}

\newcolumntype{M}[1]{>{\centering\arraybackslash}m{#1}}

\newcommand{\e}{\mathrm{e}}

\DeclareMathAlphabet{\bi}{OML}{cmm}{b}{it}

\def\be{\begin{equation}}
\def\ee{\end{equation}}
\def\bearr{\begin{eqnarray}}
\def\eearr{\end{eqnarray}}
\def\la{\langle}
\def\ra{\rangle}

\usepackage{tikz}

\begin{document}
\title{RKKY interaction in Mn-doped 4 $\times$ 4 Luttinger systems}
\bigskip
\author{Sonu Verma$^{*}$, Arijit Kundu and Tarun Kanti Ghosh\\
\normalsize
Department of Physics, Indian Institute of Technology-Kanpur,
Kanpur-208 016, India\\
$^{*}$Author to whom correspondence should be addressed: sonuv@iitk.ac.in}
\date{\today}

\begin{abstract}
We consider Mn-doped bulk zinc-blende 
semiconductors described by the 4 $ \times$ 4  Luttinger Hamiltonian. 
In these semiconductors, Mn atom acts as an acceptor providing the system a mobile hole, 
and also acts like a magnetic impurity of spin $S=5/2$.
We obtain exact analytical expressions of the hole mediated Ruderman-Kittel-Kasuya-Yoshida (RKKY) 
exchange interaction between two Mn${}^{2+}$ ions. The RKKY interaction 
of the Luttinger system consists of collinear Heisenberg-like and Ising-like interactions. 
The characteristic beating patterns appear in the range functions of the RKKY interaction owing to 
the presence of multiple Fermi wave-vectors of the underlying $j=3/2$ states. 
As an application of the analytical form of the range function, from the finite temperature 
evaluation of the correlation functions,  we calculate the contribution of RKKY 
interaction to the Curie-Weiss temperatures of a particular dilute magnetic 
semiconductor ZnMnTe where $4\times4$ Luttinger Hamiltonian is valid. 
\end{abstract}

\maketitle
\section{Introduction}
An electric control of the spin degree of freedom of a charge carrier 
is one of the primary objectives in spintronics and quantum information 
processing. The inherent spin-orbit interaction (SOI) arises due to the relativistic 
effect, which can be controlled by the spatial inversion symmetry breaking external 
electric field. The SOIs in materials give rise to many exotic phenomena. 
For example, the intrinsic spin Hall effect (SHE) arises solely due to
the spin-orbit coupling even in absence of any magnetic impurities.
After the theoretical proposal of intrinsic SHE \cite{murakami1} in $p$-doped 
III-V semiconductors described by the Luttinger Hamiltonian \cite{Lutin} for the 
spin-3/2 valence band, there is a resurgent research interest on various properties 
of the Luttinger Hamiltonian \cite{Lutin}.
This exotic phenomena has been realized experimentally in bulk n-doped semiconductors 
such as GaAs and InGaAs \cite{she-exp} as well as in two-dimensional hole gas \cite{she-exp1}. 
The hole gas is preferred over electron gas in the study of spin-related phenomena.
This is because the $p$-orbital states of the hole wave function reduces the
contact hyperfine interaction \cite{spin-coh,spin-coh1}. 
This in turn enhances the spin coherence time of
the hole charge carrier \cite{spin-coh-exp,spin-coh-exp1}. 
A large number of theoretical studies e.g. spin Hall conductivity \cite{murakami2}, 
wave packet dynamics \cite{zb1,zb2,zb3}, Hartree-Fock analysis \cite{jSc1},  
beating pattern in Friedel oscillations \cite{jSc2,jSc3}, magnetotransport 
coefficients \cite{sdh}, electrical and optical conductivities \cite{opt} of 
the Luttinger Hamiltonian have been carried out in recent past studies.

The mechanism of interaction between two localized magnetic impurities
in spintronics materials attract considerable attention.  
The RKKY interaction \cite{Ruderman,kasuya,yosida} is an indirect 
exchange interaction between two magnetic impurities mediated by 
mobile charge carriers. 
This long-range spin-spin interaction plays a
crucial role in magnetic ordering (ferromagnetic/antiferromagnetic) of the impurities 
and may help to understand the magnetic properties of the host system.
The nature of the mobile carriers (e.g. helicity, energy dispersion, 
spinor structure etc) determine the characteristics of the RKKY interaction. 
The role of the RKKY interaction in other condensed matter systems
has also been studied extensively.  
For instance, magnetoresistance in multilayer structures \cite{multilayer}, 
topological states and Majorana fermions \cite{Majorana}.
The Rashba spin-orbit coupling effect on RKKY interaction 
has been rigorously studied in 1D \cite{rkky1,rkky1-egger,rkky1-lin,rkky1-loss},
2D \cite{rkky2,rkky2-loss,rkky2-kern,multi-graphene,sonu2018,Firoz-phosp} 
and 3D \cite{rkky3} electron systems.
The strength of the range functions characterizing RKKY interaction in various systems 
oscillate with the distance between two magnetic impurities $(R)$ and decays
asymptotically as $1/R^\eta$ with $\eta $ being the
system dependent exponent.
The oscillation frequency (in units of the distance $R$) is determined by the 
density and effective mass of the charge carriers and other material parameters.

The ferromagnetic ordering in Mn-doped zinc-blende semiconductors
was first realized by Muneketa et al \cite{1st-exp}.
Subsequently it has been established that 
Mn atom is the source of local magnetic moments and also provides mobile holes
in many Mn-doped zinc-blende semiconductors (such as GaAs, GaP and ZnTe) and
show up the Curie-Weiss temperatures 
from few kelvins to few hundred kelvins 
\cite{110,fms-e,fms-e1,fms-e2,fms-e3,fms-e4,exp-theory,fms-rev,fms-e5,GaP1,GaP2}.

There have been extensive theoretical studies of ferromagnetism in Mn-doped 
zinc blende semiconductors \cite{fms-rev}, which estimate the Curie-Weiss transition 
temperature closed to the experimental findings.
The Ginzburg-Landau theory \cite{GL} has been used to describe ferromagnetic
properties of Mn-doped semiconductors.
A simple model in the low-Mn density regime was proposed \cite{hop},
in which holes are allowed to hop to the magnetic impurity sites and
interact with the magnetic moments via phenomenological
exchange interactions.
There are other models based on a polaronic picture
where a cloud of Mn-spins are polarized by a single hole \cite{polaronic}.
The concept of magnetic percolation picture was introduced to 
estimate the observed Curie-Weiss temperature \cite{percolation}. 
The ferromagnetic semiconductors have also been studied theoretically using
 the ${\bf k} \cdot {\bf p}$ kinetic-exchange effective Hamiltonian
\cite{dietl-6x6,dietl0}. There have  been several studies to explain the ferromagnetism in 
zinc-blende semiconductors using the RKKY exchange interaction \cite{diet,exp-theory,avinash,timm}.


An analytical study of RKKY interaction mediated by the states of underlying Luttinger 
Hamiltonian is still lacking. In this work, we provide an exact analytical expression 
of the RKKY exchange interaction between two magnetic Mn${}^{2+}$ ions in 3D hole-gas 
(3DHG) that follows the $ 4 \times 4 $ Luttinger 
Hamiltonian. 
It's form is different from that of 3DEG owing to the multiband 
nature of the Luttinger system. Our result displays the explicit dependence on 
the relevant band structure parameters of the Luttinger Hamiltonian. 
Using the analytical expression of the range function, we determine 
the contribution of RKKY interaction to the Curie-Weiss temperatures $T_C$ for a 
ferromagnetic semiconductor ZnMnTe.

The remainder of this paper is organized as follows.
In section II, we briefly describe the Kohn-Luttinger Hamiltonian and
its basic ground state properties. In section III,  we derive an analytical expression 
of the RKKY interaction in Mn-doped zinc blende semiconductors described by
$4\times 4$ Kohn-Luttinger Hamiltonian. 
In section IV, we compute $T_C$ for ZnMnTe and 
summarize our results in section V.

\section{Basic information}
\begin{figure}[t]
\begin{center}
\includegraphics[width=0.4\textwidth]{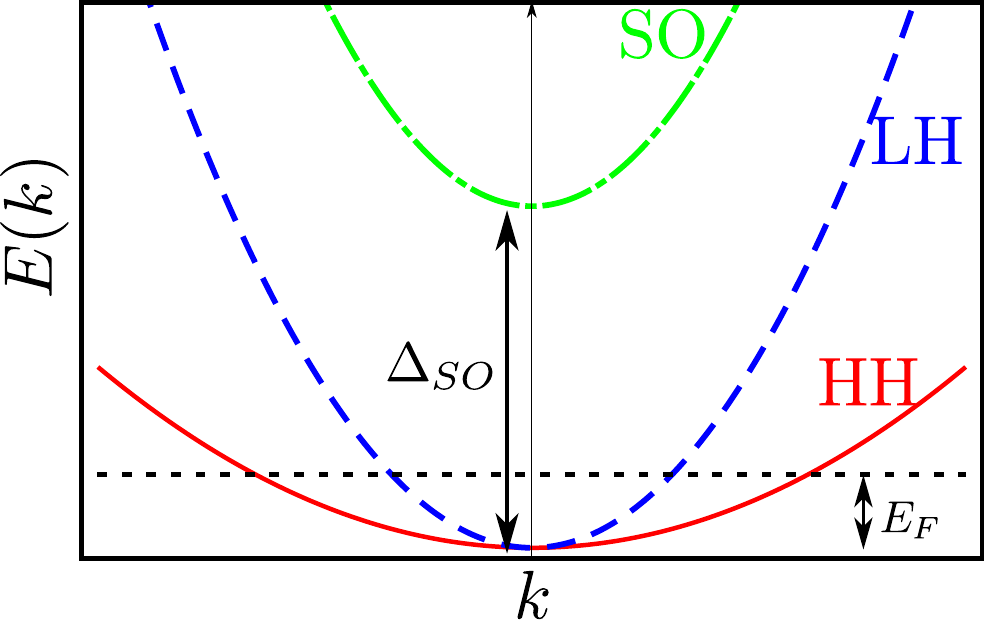}
\caption{ Sketch of energy dispersion of heavy hole, light hole and split-off bands 
for $p$-doped zinc-blende semiconductors.}\label{fig:band} 
	\end{center}
\end{figure}

The valence bands of zinc-blende semiconductors can be faithfully
described by the $6 \times 6$ Kohn-Luttinger Hamiltonian \cite{Lutin6x6} 
in the basis of total angular momentum eigenstates $|j,m_j\ra$:
$|3/2,+3/2\ra$, $|3/2,+1/2\ra$, $|3/2,-1/2\ra$, $|3/2,-3/2\ra$, 
$|1/2,+1/2\ra$, $|1/2,-1/2\ra$ as
given by
\begin{widetext}
\begin{eqnarray}\label{6by6}
H=\begin{bmatrix}
-P-Q & L & -M & 0 & \frac{1}{\sqrt{2}}L & \sqrt{2} M\\
L^\dagger & -P+Q & 0 & M & \sqrt{2}Q & -\sqrt{\frac{3}{2}} L\\
M^\dagger & 0 & -P+Q & -L & -\sqrt{\frac{3}{2}}L^\dagger & -\sqrt{2} Q\\
0 & M^\dagger & -L^\dagger & -P-Q & - \sqrt{2}M^\dagger &
\frac{1}{\sqrt{2}} L^\dagger\\
\frac{1}{\sqrt{2}}L^\dagger & \sqrt{2}Q^\dagger &
-\sqrt{\frac{3}{2}}L & -\sqrt{2}M & - P - \Delta_{\rm SO}  & 0\\
\sqrt{2}M^\dagger & -\sqrt{\frac{3}{2}}L^\dagger &
-\sqrt{2}Q^\dagger& \frac{1}{\sqrt{2}}L & 0 & - P - \Delta_{\rm SO}
\end{bmatrix},
\end{eqnarray}
\end{widetext}
where
$ P=\frac{\gamma_1 \hbar^2 k^2}{2m_0}$,
$Q = \frac{\gamma_2 \hbar^2 k^2}{2m_0} (1-3 \cos^2\theta)$,
$L = \frac{\sqrt{3}\gamma_3 \hbar^2 k^2}{2m_0}
\sin 2\theta \e^{-i \phi} $
and
$ M = \frac{\sqrt{3} \hbar^2 k^2}{2m_0} \sin^2\theta
[\gamma_2 \cos 2\phi - i \gamma_3 \sin 2\phi] $
with $(k,\theta,\phi)$ are the spherical polar coordinates of the wave
vector ${\bf k}$.
Here $m_0$ and $\Delta_{\rm SO} $ being the bare electron mass and 
the split-off energy, respectively. The dimensionless Luttinger parameters 
$\gamma_1$, $\gamma_2$ and $\gamma_3$ characterize the valence band
of the specific semiconductors.
The information about the spin-orbit coupling is contained in the 
parameters $\gamma_2$ and $\gamma_3$.
Typical band structure of the Kohn-Luttinger Hamiltonian
is shown in Fig.~\ref{fig:band}.
We can safely ignore the split-off band if the Fermi energy $(E_F)$ is
less than the split-off energy. Hence the upper-left
$4 \times 4$ matrix block in Eq. (\ref{6by6}) describes the two upper most
valence bands, usually known as heavy hole and light hole bands, approximately.

Within the spherical approximation \cite{spherical}, replacing $\gamma_2 $ 
and $\gamma_3$ by the average value $\gamma_s = 2\gamma_2/5 + 3 \gamma_3/5$,  
the $4 \times 4$ Luttinger Hamiltonian \cite{Lutin}
describing heavy-hole and light-hole states is
\begin{eqnarray}
H_0 = \frac{1}{2 m_0}\Big[ \Big(\gamma_1 + \frac{5}{2}\gamma_s \Big)\textbf{p}^{2}- 
2 \gamma_s\big(\textbf{p} \cdot \textbf{J}\big)^2\Big].
\end{eqnarray}
Here 
$m_0$ is the bare electron mass and $\textbf{J}$ is the 
spin-$3/2$ matrix operator. 
The components of the matrix operator are given by,
\begin{equation}
J_x = \begin{pmatrix}
    0 & \frac{\sqrt{3}}{2} &0 & 0  \\
    \frac{\sqrt{3}}{2} & 0 & 1 & 0 \\
    0 & 1 & 0 & \frac{\sqrt{3}}{2} \\
    0 & 0 & \frac{\sqrt{3}}{2} & 0 
  \end{pmatrix},
\end{equation}
\begin{equation}
J_y = i \begin{pmatrix}
    0 & -\frac{\sqrt{3}}{2} &0 & 0  \\
    \frac{\sqrt{3}}{2} & 0 & -1 & 0 \\
    0 & 1 & 0 & -\frac{\sqrt{3}}{2} \\
    0 & 0 & \frac{\sqrt{3}}{2} & 0 
  \end{pmatrix},
\end{equation}
\begin{equation}
J_z = \begin{pmatrix}
    \frac{3}{2} & 0 &0 & 0  \\
    0 & \frac{1}{2} & 0 & 0 \\
    0 & 0 & -\frac{1}{2} & 0 \\
    0 & 0 & 0 & -\frac{3}{2}
  \end{pmatrix}.
\end{equation}

\begin{center}
\begin{table}[h]
\def\arraystretch{2.0}
\begin{tabular}{ | c | c | c | c | c | c |c|c|}
\hline
Compound& $a$ (\r{A}) & $\gamma_1$ & $\gamma_2$ &$\gamma_3$ 
& $\gamma_s$ &  $ \Delta_{SO} $ (eV) & $ E_F $ (eV) \\ \hline
ZnTe & 6.10  &3.8 & 0.72 &1.13 & 1.068 & 0.96 & 0.1195 \\ \hline
GaAs & 5.65  & 6.98 & 2.06 & 2.93&2.58 & 0.32-0.36  & 0.1377  \\ \hline
GaP & 5.45  & 4.05 & 0.49&2.93 & 1.95 & 0.03-0.13 & 0.01116 \\
\hline
\end{tabular}
\caption{Values of various parameters for the effective Luttinger Hamiltonian 
of families of zinc-blende semiconductors, from Ref. \cite{dietl-6x6,dietl0,review-jap}.
Typical hole density in these semiconductors varies between ($10^{24}-10^{26}$) m$^{-3}$. 
Here, we have taken the hole density $n_h = 1.0\times 10^{26}$ m$^{-3}$ 
for all of them. It is clear from the table that for this typical order of density, 
the doubly degenerate two-band Hamiltonian is not applicable for GaAs and GaP , 
because $E_{F}$ is of the order of $\Delta_{SO}$. But for ZnTe, $E_{F} \ll \Delta_{SO}$ 
($\Delta_{SO}\approx 8E_{F}$), so doubly degenerate two-band Luttinger Hamiltonian will 
be a good approximation. From now onwards we will only focus on ZnTe as an example 
to elaborate our results.}
\end{table}
\end{center}
The above Hamiltonian $H_0$ is rotationally invariant and
commutes with the helicity operator $ \hat \Lambda ={\bf k}\cdot{\bf J}/k$
so that its eigenvalues $\lambda= \pm 3/2, \pm 1/2$ are good quantum numbers.
Here $\lambda=\pm 3/2$ and $\lambda= \pm 1/2$  correspond to the heavy hole
and light hole states, respectively. Therefore, the eigenstates of the helicity operator
are the same as the eigenstates of the Hamiltonian $H_0$. 
The energy dispersion of the heavy and light hole states are given by
$ E_{h/l}({\bf k}) = (\hbar k)^2/(2m_{h/l}) $ with
$ m_{h/l} = m_0/(\gamma_1 \mp 2\gamma_s) $ are the heavy and light hole masses, respectively.
The two-fold degeneracy of heavy and light hole branches is due to the
consequence of the space inversion and time-reversal symmetries of the Luttinger 
Hamiltonian. 
Using the basis of eigenstates of $J_z$,
the eigenspinors $| \lambda,{\bf k} \ra $ for $\lambda=3/2$ and $\lambda=1/2$ 
can be written as 
\begin{align}
|3/2,{\bf k} \ra &=\begin{pmatrix}\label{W32}
\cos^3\frac{\theta}{2}e^{(-3i/2)\phi}\\
\sqrt{3}\cos^2\frac{\theta}{2}\sin\frac{\theta}{2} e^{(-i/2)\phi}\\
\sqrt{3}\cos\frac{\theta}{2}\sin^2\frac{\theta}{2} e^{(i/2)\phi}\\
\sin^3\frac{\theta}{2}e^{(3i/2)\phi}
\end{pmatrix}
\end{align}
and
\begin{align}\label{W12}
|1/2,{\bf k} \ra &=\begin{pmatrix}
-\sqrt{3}\cos^2\frac{\theta}{2}\sin\frac{\theta}{2}e^{(-3i/2)\phi}\\
\cos\frac{\theta}{2}\Big(\cos^2\frac{\theta}{2}-2\sin^2\frac{\theta}{2}\Big) e^{(-i/2)\phi}\\
\sin\frac{\theta}{2}\Big(2\cos^2\frac{\theta}{2}-\sin^2\frac{\theta}{2}\Big) e^{(i/2)\phi}\\
\sqrt{3}\cos\frac{\theta}{2}\sin^2\frac{\theta}{2}e^{(3i/2)\phi}
\end{pmatrix}.
\end{align}
The remaining spinors for $\lambda = -3/2$ and $\lambda = -1/2$ 
can easily be obtained from  Eq. (\ref{W32}) and Eq. (\ref{W12})
under the spatial inversion operations 
$\theta\rightarrow\pi-\theta$ and $\phi\rightarrow\pi+\phi$.

Performing the standard procedure, the Fermi energy is given by
\begin{eqnarray}\label{fermi}
E_{F} = \frac{(\hbar {k_{\rm F}^{0e}})^2}{2m_0}\bigg[\frac{\gamma_1^2 -
4\gamma_s^2}{\big[(\gamma_1-2\gamma_s)^{3/2}+(\gamma_1+2\gamma_s)^{3/2}\big]^{2/3}}\bigg]
\end{eqnarray} 
and the corresponding Fermi wave-vectors $k_F^{h/l}$ for heavy and light-hole bands, respectively,
are given by 
\begin{eqnarray} \label{FerVec}
k_{F}^{h/l}=k_{\rm F}^{0e} 
\frac{m_{h/l}^{1/2} }{(m_{h}^{3/2}+m_{l}^{3/2})^{1/3}},
\end{eqnarray}
where $k_{F}^{0e}=(3\pi^2n_h)^{1/3}$ with $n_h$ being the hole density. Also the density of states 
at Fermi energy is $\rho(E_{F})=(m_{h}^{3/2}+m_{l}^{3/2})^{2/3}k_{F}^{0e}/(\pi^2\hbar^2)$. 
For a charge carrier with spin $s$ and carrier density $n_c$, we define 
$k_{F}^{0h/0e}=(\frac{6\pi^2 n_c}{2s+1})^{1/3}$, where $s=3/2$ for hole gas and 
$s=1/2$ for electron gas, which implies $k_{F}^{0h}=k_{F}^{0e}/2^{1/3}$.
Various parameters along with the Fermi energy for three different zinc blende 
semiconductors are given in Table I.

The Green function $G(\textbf{k},\omega +i 0^{+}) = [\omega +i 0^{+}-H_0]^{-1}$ of 
the $4 \times 4$ Luttinger Hamiltonian is then given by \cite{opt}
\begin{widetext}
\begin{align}
G(\textbf{k},\omega)&=\sum_{\lambda}\frac{1}{E_{\lambda}- \omega - i0^{+}}\begin{pmatrix}
-\frac{1}{2}-\frac{\lambda(1+3\cos (2\theta))}{8}& - 
\frac{\lambda\sqrt{3}}{4}\sin (2\theta) e^{-i\phi} & - 
\frac{\lambda\sqrt{3}}{4}\sin^{2} \theta e^{-2i\phi} & 0  \\
-\frac{\lambda\sqrt{3}}{4}\sin (2\theta) e^{i\phi} & -\frac{1}{2} + 
\frac{\lambda(1+3\cos (2\theta))}{8} & 0 & - 
\frac{\lambda\sqrt{3}}{4}\sin^2 \theta e^{-2i\phi} \\
- \frac{\lambda\sqrt{3}}{4}\sin^2 \theta e^{2i\phi} & 0 & - 
\frac{1}{2}+\frac{\lambda(1+3\cos (2\theta))}{8} &  
\frac{\lambda\sqrt{3}}{4}\sin (2\theta) e^{-i\phi} \\
0 & -\frac{\lambda\sqrt{3}}{4}\sin^2\theta e^{2i\phi} & 
\frac{\lambda\sqrt{3}}{4}\sin (2\theta) e^{i\phi} & -\frac{1}{2} - 
\frac{\lambda(1+3\cos (2\theta))}{8}
\end{pmatrix},\label{eq:G}
\end{align}
\end{widetext}
which we will use in the next section. 

In these semiconductors, the magnetic impurities (Mn$^{2+}$ state having localised $d$-orbitals) 
interact with each other by valance holes (having $p$-orbital) through the exchange interaction.
So we assume $p$-$d$ type contact exchange interaction between the hole spin 
$\mathbf{J}(\mathbf{r})$ and the spins of the magnetic impurities $\mathbf{S}_j$, 
at positions $\mathbf{R}_j$ as 
\begin{eqnarray}
H_{p-d} =  J_{pd}^{*} \sum_{j=1,2} \textbf{S}_{j} \cdot \textbf{J}(\mathbf{r}) \hspace{1.5mm}
\delta(\textbf{r}-\textbf{R}_{j})
\end{eqnarray}
with $J_{pd}^{*}$ is the strength of $p$-$d$ exchange interaction, whose dimension
is energy times volume. Here $J_{pd}^{*}=J_{pd}/3$, with $J_{pd}=50$ eV $\r{A}^3$ for Mn doped ZnTe dilute magnetic semiconductors\cite{dietl0}.
Thus the total Hamiltonian of the system is
$ H = H_0 + H_{p-d}$. By considering $H_{p-d}$ as a perturbation, 
the RKKY interaction is the second order correction  to the ground state energy of $H_0$.

\begin{figure}[t]
\begin{center}
\includegraphics[width=0.49\textwidth]{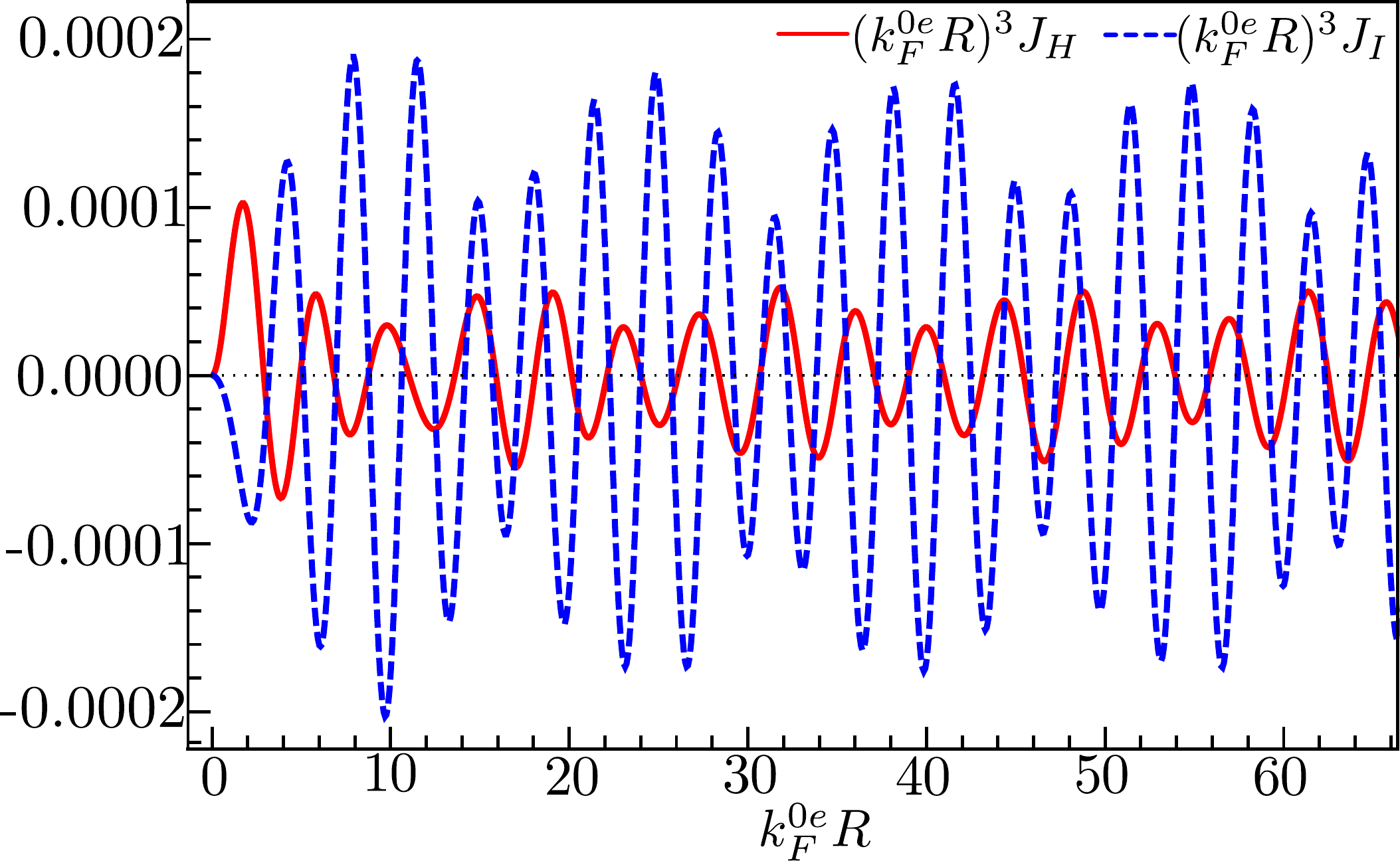}
\caption{The range functions of RKKY interaction, from Eqs.~\ref{eq:rkky1} and \ref{eq:rkky2}, 
for Mn-doped ZnTe are plotted to show the beating pattern.}\label{fig:RKKY}
\end{center}
\end{figure}

\section{RKKY interaction}
The RKKY interaction between two impurity spins $\mathbf{S}_1$ and $\mathbf{S}_2$ put at a 
distance $\mathbf{R}$, at zero temperature, can be computed using second-order 
perturbation theory and it is expressed as follows
\begin{eqnarray} \label{rkky}
H_{{\rm RKKY}}& = & - \frac{J_{pd}^{2} }{9\pi}{\rm Im}\int_{-\infty}^{E_{F}}d\omega\hspace{1mm} 
{\rm Tr}[ (\textbf{S}_{1} \cdot \textbf{J}) \hspace{1mm}G(\textbf{R},\omega +i 0^{+}) 
\nonumber \\ 
& \times &(\textbf{S}_{2} \cdot \textbf{J})\hspace{1mm}G(-\textbf{R},\omega +i 0^{+})],
\end{eqnarray}
where $ {\rm Tr} $ indicates a trace over the spin degree of freedom. 
The energy-coordinate representation of the Green's function is given by
\begin{eqnarray}
G(\pm\textbf{R},\omega +i 0^{+})=\int \frac{d^3 k}{(2\pi)^3}
G(\textbf{k},\omega +i 0^{+}) e^{\pm i \textbf{k} \cdot \textbf{R}},
\end{eqnarray}
where we use the Green's function as in Eq.~(\ref{eq:G}). Without loss of generality, 
we consider the spins along the $z$ axis (i.e, $\mathbf{R} = R\hat{z}$) and after 
some tedious algebra, we obtain the RKKY interaction, containing only the Heisenberg 
and the Ising terms as
\begin{eqnarray}
H_{\rm RKKY} & = & \label{rkky-range}
J_{H}\textbf{S}_{1}.\textbf{S}_{2}+ J_{I}S_{1z} S_{2z},
\end{eqnarray}
where $ J_{H}$ and $J_{I}$ denote range functions for the collinear Heisenberg 
and Ising terms, respectively.
The non-collinear Dzyaloshinsky-Moriya (DM) \cite{dm1,dm2} coupling term is absent since the Luttinger 
Hamiltonian is invariant under spatial inversion.
The detail derivation of $J_H$ and $J_I$ are given in the appendix, with the final form being
\begin{widetext}
	\begin{align}
	J_H & = 
	- \frac{J_{pd}^{2} }{(2\pi)^3} \Big(\frac{ m_h}{9\hbar^2 R^4}\Big)
	\Big[
	-\frac{9 \cos (2 \zeta^{h}_{F})}{2\zeta^{h}_{F}}
	-\Big( \frac{9\delta^2}{2 \zeta^{l}_{F}} + 2 \delta^2 \zeta^{l}_{F} \Big)
	\cos (2 \zeta^{l}_{F})
	+ \Big(-\frac{6 \delta^2 \zeta^{h}_{F} }{(1+\delta)} + \frac{9(1+\delta)}{ 2\zeta^{h}_{F} }\Big)
	\cos(\zeta^{h}_{F}+ \zeta^{l}_{F} ) \nonumber \\
	&+
	\Big(- \frac{9}{2} + \frac{9}{4 (\zeta^{h}_{F})^2} \Big) \sin (2 \zeta^{h}_{F})
	+ \Big( \frac{5 \delta^2}{2} + \frac{9 \delta^2}{4 (\zeta^{l}_{F})^2} \Big) \sin (2 \zeta^{l}_{F})
	- \Big(\frac{9}{2 (\zeta^{h}_{F})^2} +
	\frac{(3\delta^3 - 12 \delta^2 - 9 \delta)}{(1+\delta)^2} \Big) \sin (\zeta^{h}_{F}+ \zeta^{l}_{F})
	\nonumber \\
	&+
	\frac{9}{2}{\rm si}(2 \zeta^{h}_{F}) - \frac{15\delta^2}{2}{\rm si}(2 \zeta^{l}_{F}) +
	\Big(\frac{15\delta^2}{ 2  } - \frac{9}{2} \Big){\rm si}( \zeta^{h}_{F}+\zeta^{l}_{F})
	\Big],\label{eq:rkky1}\\
	J_{I} & = 
	- \frac{J_{pd}^{2} }{(2\pi)^3} \Big(\frac{3 m_h}{18\hbar^2 R^4}\Big)
	\Big[\Big(\frac{18}{\zeta_{F}^h} - 3 \zeta_{F}^h \Big) \cos(2 \zeta^{h}_{F})
	+ \Big(\frac{18 \delta^2}{\zeta_{F}^l} + \delta^2 \zeta_{F}^l \Big) \cos(2 \zeta^{l}_{F})
	+ \Big(\frac{4 \delta^2 \zeta_{F}^h}{1+\delta} - \frac{18(1+\delta)}{\zeta_{F}^h} \Big)
	\cos(\zeta_{F}^h + \zeta_{F}^l) \nonumber \\
	& + 
	\Big(\frac{27}{2} -\frac{9}{(\zeta_{F}^h)^2} \Big) \sin (2 \zeta^{h}_{F})
	- \Big(\frac{\delta^2}{2} + \frac{9 \delta^2}{(\zeta_{F}^l)^2} \Big) \sin (2 \zeta^{l}_{F})
	+ \Big( \frac{18}{(\zeta_{F}^h)^2} - \frac{4\delta(6+7\delta)}{(1+\delta)^2}\Big)
	\sin(\zeta_{F}^h + \zeta_{F}^l) \nonumber \\
	& - 
	6 {\rm si}(2 \zeta^{h}_{F})  + 18 \delta^2 {\rm si}(2 \zeta^{l}_{F})
	+ 6(1-3\delta^2) {\rm si}( \zeta^{h}_{F}+\zeta^{l}_{F})\Big],\label{eq:rkky2}
	\end{align}
\end{widetext}
where, $\zeta^{\lambda}_{F} = k^{\lambda}_F R$ and 
$\delta = \sqrt{m_l/m_h} < 1$ and ${\rm si}(x) =\int_{0}^{x}\frac{\sin t}{t} dt $ 
is the sine integral. It is evident from the above equations that the 
$J_H$ and $J_I$ oscillate with the distance between the spins, $R$, 
with multiple frequencies (due to the two different Fermi-wave vectors of the 
spin $s=3/2$ hole states), giving rise to beating pattern. 
The variation of $J_H$ and $J_I$ for a typical set of parameters has 
been plotted in Fig.~\ref{fig:RKKY}. For large distances $J_{H}$ and $J_{I}$  (upto $ 1/R^3 $ term) 
can be approximated as
\begin{widetext}
\begin{align}
J_{H}& \simeq  \frac{ J_{pd}^{2} }{(2\pi)^3} 
\frac{2 m_h \delta^2}{9\hbar^2 R^4} 
\Big[\zeta^{l}_{F} \cos (2 \zeta^{l}_{F}) +
\frac{3 \zeta^{h}_{F} }{(1+\delta)} \cos (\zeta^{h}_{F}+ \zeta^{l}_{F}) \Big],\\
J_{I}& \simeq \frac{ J_{pd}^{2} }{(2\pi)^3}
\frac{3 m_h}{ 18\hbar^2 R^4}
\Big[3 \zeta^{h}_{F} \cos (2 \zeta^{h}_{F}) - \delta^2 \zeta^{l}_{F}\cos (2 \zeta^{l}_{F}) 
+ \frac{4 \delta^2}{(1 + \delta)} \zeta^{h}_{F}
\cos (\zeta^{h}_{F}+ \zeta^{l}_{F}) \Big].
\end{align}
\end{widetext}

The nature of coupling (ferromagnetic/antiferromagnetic) between the magnetic impurities 
for a particular semiconductor is determined by the density of holes and the distance 
between the magnetic impurities.

\begin{figure}[ht]
\begin{center}
\includegraphics[width=0.49\textwidth]{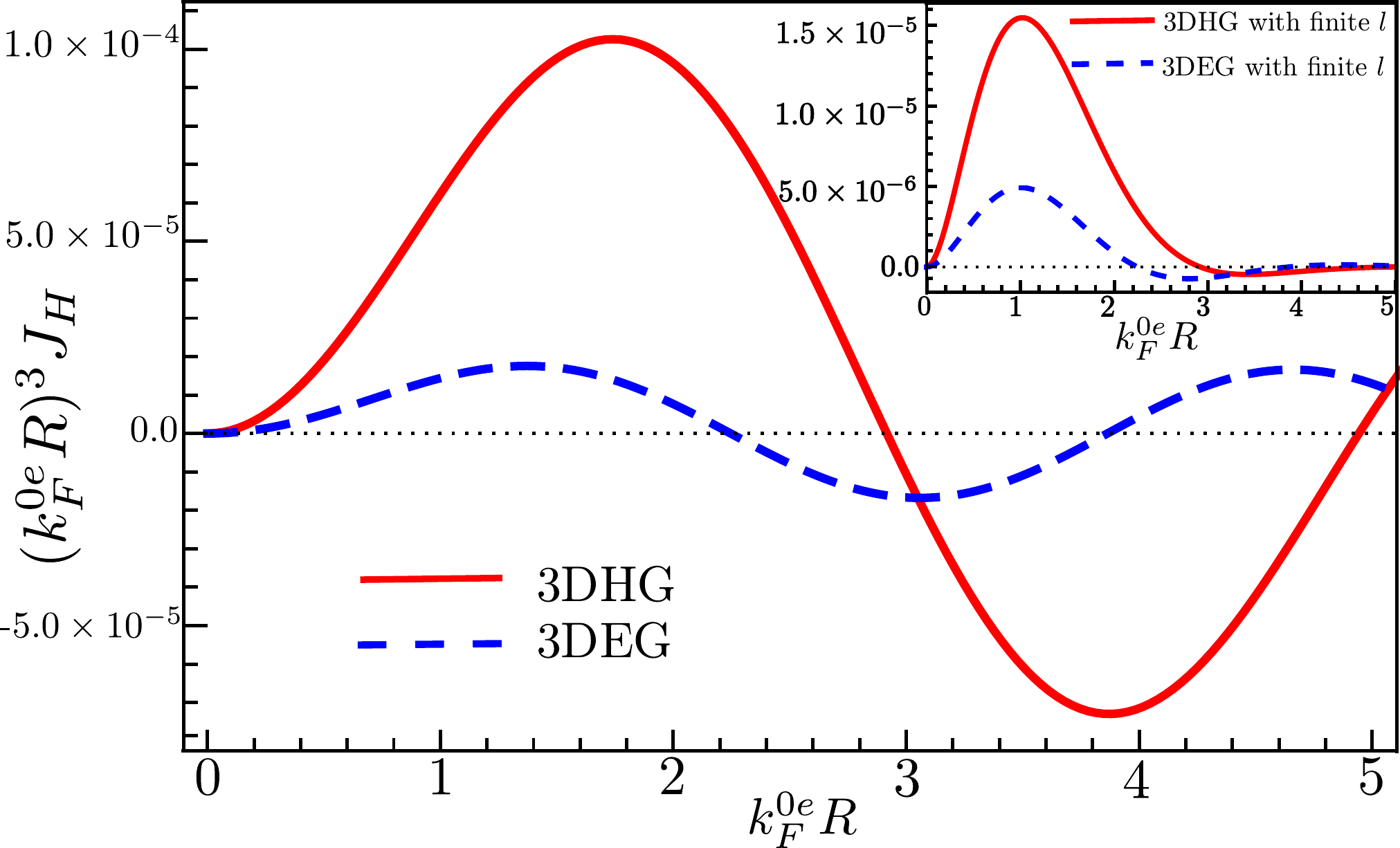}
\caption{ Comparison of the Heisenberg term of the RKKY interaction for 
Zn$_{1-x}$Mn$_x$Te, calculated using 3DEG Hamiltonian with only one band 
together with different carrier mass ($m_h=0.5m_0$) and using the Luttinger 
Hamiltonian for 3DHG with two degenerate bands together with two different masses 
$m_h$ and $m_l$. Due to finite mean free path ($l$) of carriers, the response 
function decays exponentially. So we have also plotted in the inset of this 
figure the range function for 
the Heisenberg term  with an exponential decay factor for $T_C$ calculation 
(see Eq.~(\ref{eq:Jeff})). Due to this exponential decay, we require only few 
nearest neighbour distances ($k_{F}^{0e}R\approx$ upto $2$) to compute $T_C$. 
We have taken the mean free paths $l=0.65$ nm and $l=0.5$ nm for 3DEG case and 
3DHG cases respectively.}\label{fig:comparison}
\end{center}
\end{figure}
The RKKY interaction of the hole doped semiconductors is different in nature from that
of 3DEG, where such beating pattern is absent due to single band nature. The comparison 
of the two situations is plotted in Fig.~\ref{fig:comparison}. 
It can be easily checked that the Luttinger Hamiltonian reduces to that of the conventional 
3DEG by setting $\gamma_1=1 $ and $\gamma_s = 0$. In this limit, $m_h=m_l=m_0$ and 
$ k_F^h=k_F^l= k_F^{0h}$, the Ising-like range function exactly vanishes i.e, $ J_I =0 $ and 
the Heisenberg-like range function $J_H$ looks similar to the known form of the 
conventional 3DEG case  \cite{diet,rkkyanyD},
\begin{align}
J_H &= -\frac{2J_{pd}^{2}(k_{F}^{0h})^3}{\pi}\tilde{ \chi}_{h}
\mathcal{F}(2k_{F}^{0h}R),
\end{align}
where $\mathcal{F}(y)=\frac{\sin(y)-y\cos(y)}{y^4}$ and $\tilde{\chi}_{h} = 10m_0 k_{F}^{0h}/36\pi^2\hbar^2$ is the static hole susceptibility.
We have written above expression as a function of $k_{F}^{0h}$, because in 
the limiting situation $m_{l}=m_{h}$, we will have four degenerate bands.

\section{Mean-field Curie-Weiss temperature}
In this section, we will compute the contribution of the RKKY interaction to the 
Curie-Weiss temperature. In the mean-field approximation, the 
Curie-Weiss temperature $T_c$ is given by
\begin{eqnarray}\label{eq:Tc}
k_{B}T_C = \frac{x}{3}S(S+1){ J_{H}^{\rm eff}},
\end{eqnarray}
where $x$ is the magnetic dopant concentration, $S=5/2$ (for magnetic impurity Mn$^{+2}$ state) 
and spatial average of the Heisenberg-type response $ J_{H}^{\rm eff}$ is given by
\begin{eqnarray}\label{eq:Jeff}
{J_{H}^{\rm eff}}= \sum_r z_r  J_{H}(r_{ij})e^{-r_{i j}/l}.
\end{eqnarray}
Here $z_r$ is the number of $r$-th neighbors (Ga-Ga), $l$ is carrier transport 
mean-free path which is introduced as an exponential factor in the effective interaction 
to take into account the fact that carriers can not pass the information 
from one magnetic impurity to other magnetic impurities which are situated at 
distances greater than the mean-free path of the carriers.
We have used upto fourth nearest neighbour distances in the above summation, which
is sufficient for Mn-doped ZnTe with mean-free path $l=0.5$ nm.

The origin of ferromagnetism in some dilute magnetic 
semiconductors is mostly due to the RKKY interaction mediated by valence 
holes. For these semiconductors, one can calculate $T_C$, 
using the given analytical form of $J_H$ with appropriate material parameters 
like mean free path of itinerant holes and strength of indirect exchange 
interaction $J_{pd}$, provided the density of holes is such that 
$E_F \ll \Delta_{SO}$. For higher hole densities, two-band approximation 
will not be valid and one has to take into account the contribution from the 
split-off band. Then we have to work with $6 \times 6$ Kohn-Luttinger Hamiltonian, 
which will be more complicated. As mentioned in the Table I, for dilute magnetic 
semiconductors like (Ga,Mn)As, (Ga,Mn)P, our results will not be valid because for 
these systems hole density is such that Fermi energy becomes comparable with the 
split-off energy and then there will also be contribution from this band 
which is not included in our calculation. 
The critical temperature $T_C$ may also depend on other effects like
hole-hole interaction, super-exchange etc \cite{dietl0}.
For just an application of our analytical result, we present in Table II, 
the calculated $T_C$ of Mn-doped ZnTe dilute magnetic semiconductor 
for different magnetic dopant concentrations. 
For this material, we have taken mean free path \cite{exp-theory} of the carrier holes 
$l= 0.5 $ nm and the strength of the exchange interaction 
\cite{exp-theory,dietl-6x6,dietl0} $J_{pd}^{*} =  50/3$  ${\rm eV \r{A}}^3$.

The RKKY interaction through the underlying Luttinger system is markedly different 
from a 3DEG system, although the resultant $T_C$ computed from either system may 
match closely. This is because of the small mean free path of the relevant 
semiconductors. The long range features such as the beating pattern do not have a 
strong contribution in $T_C$ as the range function is now multiplied by an exponentially 
decaying function, as shown in Fig.~\ref{fig:comparison}. 
This is the reason for successful 
prediction of the Curie-Weiss temperature in previous studies based 
in 3DEG modeling of these systems.

\begin{center}
\begin{table}
\def\arraystretch{1.5}
\begin{tabular}{ |c |c| }
\hline Mn-fraction & \multicolumn{1}{|c|}{$T_C$}\\
\hline
($x$) & (in Kelvin)  \\
\hline
0.015 & 0.48 \\
\hline
0.022 & 0.70 \\
\hline
0.043 & 1.37 \\
\hline
0.053 & 1.69  \\
\hline
0.071 & 2.27 \\
\hline
\end{tabular}
\caption{Curie-Weiss temperature $T_C$ of dilute magnetic
semiconductor (Zn,Mn)Te for various Mn-fraction $x$, which is calculated using Eq.~(\ref{eq:Jeff}). 
Experimental values of $T_C$: 
for Zn$_{1-x}$Mn$_{x}$Te, $T_C\approx (1-10) $ K, for $x=(0.01-0.05)$ 
\cite{dietl-6x6,dietl0}. }
\end{table}
\end{center}

\begin{figure}[t]
\begin{center}
\includegraphics[width=0.45\textwidth]{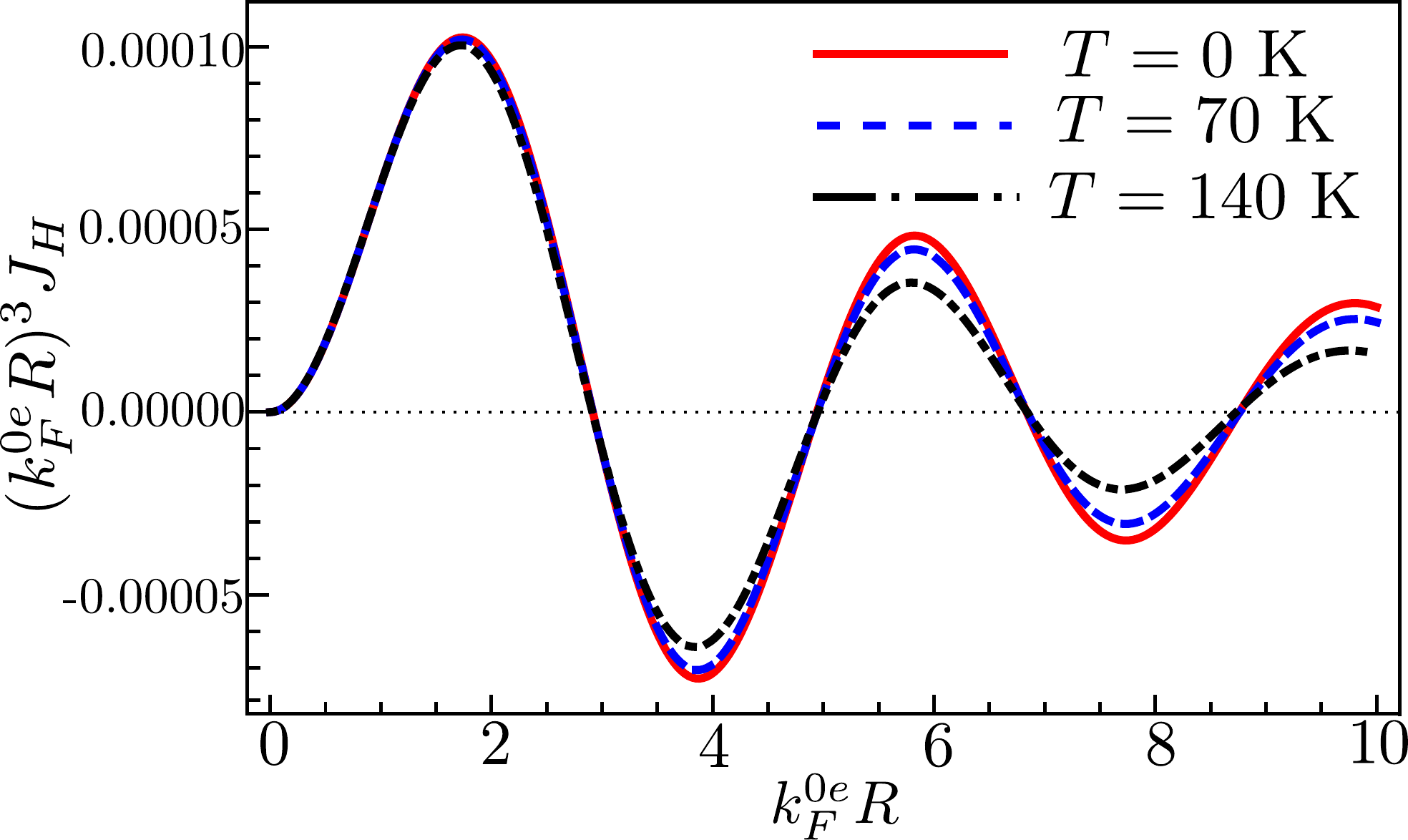}
\caption{Temperature dependence of the RKKY interaction strength for Mn-doped ZnTe, showing 
the change in $J_H$ is small with the temperature for 
the distance between the impurities contributing to the mean-field $T_C$ calculation. }\label{fig:temp}
\end{center}
\end{figure}

We have also computed the transition temperature $T_C$ using the temperature dependence 
of the RKKY interaction \cite{Maldague} given by
\begin{align*}
J_{H}(E_F, R, T)=\int_{-\infty}^{\infty} 
\frac{J_{H}(\omega, R)}{4 k_{B}T \cosh^{2}((E_F-\omega)/2 k_{B}T)}d\omega.
\end{align*}
For the relevant parameters, $J_{H}(E_F, R, T)$ depends weakly on the temperature, 
as shown in Fig.~\ref{fig:temp}, for $R$ is of the order of the mean free path. 
Considering the spatial average $J_{H}^{\rm eff}(T)$, defined in equation 
Eq.~(\ref{eq:Jeff}), one solves Eq.~\ref{eq:Tc} self-consistently to find 
the transition temperature $T_c$. As $J_H$ depends weakly on temperature for $R$ is of 
the order of mean free path, 
the resulting finite temperature estimation of $T_C$ does not differ significantly 
from what is presented in Table II.

\section{Summary and conclusions}
In this work, we have studied RKKY interaction in Mn-doped bulk zinc-blende semiconductors, 
described by the $4\times4$ Luttinger Hamiltonian. The analytical form of the interaction, 
as in Eqs.~(\ref{eq:rkky1}) and (\ref{eq:rkky2}), describes the effect of the multiple 
bands through the presence of multiple frequency of oscillation giving rise to 
beating pattern in the range functions. 
For systems which are described by $4\times4$ Luttinger Hamiltonian, 
we calculate the contribution of the RKKY interaction to the ferromagnetic transition 
temperature $T_C$.
As an application of our analytical result, we have calculated $T_C$  for a zinc-blende 
semiconductor (Zn,Mn)Te. We also found that the spin-spin correlation is insensitive to temperature 
for small distances of the order of mean free path. Therefore finite temperature 
estimation of $T_C$  does not differ significantly from the estimation at 
$T=0$ K.

\section{Acknowledgement}
We would like to thank Tomasz Dietl and Joydeep Chakraborthy for useful discussion. 

\begin{widetext}

\appendix

\section{Derivation of RKKY interaction}

In this section we provide detail derivation of the RKKY interaction.
For convenience, we choose $ {\bf R} = R \hat{z}$ with $ R = |{\bf R}_1 - {\bf R}_2|$.
It should be noted here that our results are the same for any arbitrary direction of ${\bf R} $.
The Green's function $ G(\pm \textbf{R},\omega +i 0^{+})$ is reduced to 
the following diagonal matrix:
\begin{equation} \label{rkky1}
G(\pm \textbf{R},\omega +i 0^{+})
=
\begin{pmatrix}
    P_0 & 0 &0 & 0  \\
    0 & Q_0 & 0 & 0 \\
    0 & 0 & Q_0 & 0 \\
    0 & 0 & 0 & P_0
\end{pmatrix}.
\end{equation}
Here the diagonal elements $P_0$ and $Q_0$ are expressed as 
$ 4 P_0 = - [I_h^{(1)}+ 3 I_h^{(2)}+ 3 I_l^{(1)}- 3 I_l^{(2)} ]$ and
$ 4 Q_0 = - [3I_h^{(1)}- 3 I_h^{(2)}+  I_l^{(1)} + 3 I_l^{(2)}] $  
with the integrals $ I_{\lambda}^{(1)} $ and $I_{\lambda}^{(2)}$ 
are given by
\begin{eqnarray}
I^{(1)}_{\lambda} = \int \frac{d^3 k}{(2\pi)^3} 
\frac{e^{i k R \cos\theta}}{(E_{\lambda} - \omega - i0^{+})},
\end{eqnarray}
and
\begin{eqnarray}
I^{(2)}_{\lambda} = \int \frac{d^3 k}{(2\pi)^3} 
\frac{\cos^{2}\theta e^{i k R \cos\theta}}{(E_{\lambda} - \omega - i0^{+})}.
\end{eqnarray}

Considering $R>0$ and performing the three dimensional integrations, we
obtain
\begin{eqnarray}
I^{(1)}_{\lambda} & = & \frac{m_{\lambda}}{2 \pi\hbar^2 } 
\frac{e^{i k_{\lambda} R }}{R} 
\end{eqnarray}
and
\begin{eqnarray}
I^{(2)}_{\lambda} & = & 
\frac{m_{\lambda}}{\pi  \hbar^2 R} 
\Big[\frac{1}{k_{\lambda}^2 R^2} + e^{i k_{\lambda} R} 
\Big(\frac{1}{2} + \frac{i}{k_{\lambda} R} - \frac{1}{k_{\lambda}^2 R^2}\Big)\Big]   
\end{eqnarray}
with $ k_{\lambda} = \sqrt{2 m_{\lambda}(\omega + i 0^+)/\hbar^2}$.
The above expressions of $I^{(1)}_{\lambda} $ and $ I^{(2)}_{\lambda}$
remains valid for $R<0$ as well. Now the final expressions for 
the components of the Green's function are,
\begin{eqnarray} \label{P1}
P_0 & = & - \frac{m_{h}}{2\pi \hbar^2} \frac{e^{i k_{h} R}}{R}
\Big(1 + \frac{3i}{2k_h R} - \frac{3}{2k_{h}^{2} R^2}\Big) +  
\frac{3 m_{l}}{4 \pi \hbar^2} \frac{e^{i k_{l} R} }{R} 
\Big(\frac{i}{k_l R} - \frac{1}{k_{l}^{2} R^2}\Big)
\end{eqnarray}
and
\begin{eqnarray} \label{Q1}
Q_0 & = & 
\frac{3 m_{h}}{4 \pi \hbar^2} \frac{e^{i k_{h} R} }{R} 
\Big(\frac{i}{k_h R} - \frac{1}{k_{h}^{2} R^2}\Big) 
- \frac{m_{l}}{2\pi \hbar^2}\frac{e^{i k_{l} R}}{R}
\Big( 1 + \frac{3i}{2k_h R} - \frac{3}{2k_{h}^{2} R^2}\Big).
\end{eqnarray}

Using Eq. (\ref{rkky1}), Eq. (\ref{rkky}) can be reformulated as
\bearr
H_{RKKY} & = & J_H {\bf S}_1 \cdot {\bf S}_2 + J_I S_{1z}S_{2z},
\eearr
where the integral expressions of $J_H$ and $J_I$ are
\bearr
J_H & = & - \frac{4 J_{pd}^{2} }{9\pi} {\rm Im} \int_{-\infty}^{E_f} Q_0 (3P_0 + 2Q_0) d\omega \\
J_I & = & - \frac{6 J_{pd}^{2} }{9\pi} {\rm Im} \int_{-\infty}^{E_f} (3P_0^2 - Q_0^2 - 2 P_0Q_0)
d\omega.
\eearr

Using Eqs. (\ref{P1}) and (\ref{Q1}) into tne integral expressions of the 
range functions $J_H$ and $J_I$ and performing the energy integral, 
the exact analytical expressions of the range functions are 
\begin{eqnarray} \label{J_H}
J_H & = &
- \frac{J_{pd}^{2}}{(2\pi)^3} \Big(\frac{  m_h}{9\hbar^2 R^4}\Big)
\Big[
-\frac{9 \cos (2 \zeta^{h}_{F})}{2\zeta^{h}_{F}}
-\Big( \frac{9\delta^2}{2 \zeta^{l}_{F}} + 2 \delta^2 \zeta^{l}_{F} \Big)
\cos (2 \zeta^{l}_{F})
+ \Big(-\frac{6 \delta^2 \zeta^{h}_{F} }{(1+\delta)} + \frac{9(1+\delta)}{ 2\zeta^{h}_{F} }\Big)
\cos(\zeta^{h}_{F}+ \zeta^{l}_{F} ) \nonumber \\
&+&
\Big(- \frac{9}{2} + \frac{9}{4 (\zeta^{h}_{F})^2} \Big) \sin (2 \zeta^{h}_{F})
+ \Big( \frac{5 \delta^2}{2} + \frac{9 \delta^2}{4 (\zeta^{l}_{F})^2} \Big) \sin (2 \zeta^{l}_{F})
- \Big(\frac{9}{2 (\zeta^{h}_{F})^2} +
\frac{(3\delta^3 - 12 \delta^2 - 9 \delta)}{(1+\delta)^2} \Big) \sin (\zeta^{h}_{F}+ \zeta^{l}_{F})
\nonumber \\
&+&
\frac{9}{2}{\rm si}(2 \zeta^{h}_{F}) - \frac{15\delta^2}{2}{\rm si}(2 \zeta^{l}_{F}) +
\Big(\frac{15\delta^2}{ 2   } - \frac{9}{2} \Big){\rm si}( \zeta^{h}_{F}+\zeta^{l}_{F})
\Big],
\end{eqnarray}
\begin{eqnarray}
J_I & = & \label{J_I}
- \frac{J_{pd}^{2} }{(2\pi)^3} \Big(\frac{3 m_h}{18\hbar^2 R^4}\Big)
\Big[\Big(\frac{18}{\zeta_{F}^h} - 3 \zeta_{F}^h \Big) \cos(2 \zeta^{h}_{F})
+ \Big(\frac{18 \delta^2}{\zeta_{F}^l} + \delta^2 \zeta_{F}^l \Big) \cos(2 \zeta^{l}_{F})
+ \Big(\frac{4 \delta^2 \zeta_{F}^h}{1+\delta} - \frac{18(1+\delta)}{\zeta_{F}^h} \Big)
\cos(\zeta_{F}^h + \zeta_{F}^l) \nonumber \\
& + &
\Big(\frac{27}{2} -\frac{9}{(\zeta_{F}^h)^2} \Big) \sin (2 \zeta^{h}_{F})
- \Big(\frac{\delta^2}{2} + \frac{9 \delta^2}{(\zeta_{F}^l)^2} \Big) \sin (2 \zeta^{l}_{F})
+ \Big( \frac{18}{(\zeta_{F}^h)^2} - \frac{4\delta(6+7\delta)}{(1+\delta)^2}\Big)
\sin(\zeta_{F}^h + \zeta_{F}^l) \nonumber \\
& - &
6 {\rm si}(2 \zeta^{h}_{F})  + 18 \delta^2 {\rm si}(2 \zeta^{l}_{F})
+ 6(1-3\delta^2) {\rm si}( \zeta^{h}_{F}+\zeta^{l}_{F})\Big].
\end{eqnarray}
Here, 
${\rm si}(x) = \int_{0}^{x}\frac{\sin t}{t} dt $ is the sine integral.
\section{Derivation of static spin susceptibility}
Following Ref.\cite{dietl0},  the longitudinal static spin susceptibility for this system can be written as
\begin{eqnarray}\label{sss}
\chi({\bf q},\mu,T) = \frac{1}{9V}\sum_{\lambda_1,\lambda_2,{\bf k}} 
|\la\lambda_1,{\bf k}|J_{z}|\lambda_2,{\bf k} + {\bf q} \ra|^2 
\frac{f(\lambda_1,{\bf k}) - f(\lambda_2,{\bf k} + {\bf q})}{E_{\lambda_2}
	({\bf k} + {\bf q}) - E_{\lambda_1}({\bf k})},
\end{eqnarray}
where  $|\lambda,{\bf k} \ra$ is the eigen spinor of the Hamiltonian with eigen energy 
$E_{\lambda}({\bf k})$ defined in the main text, 
$f(\lambda,{\bf k}) = 1/(\exp(\beta(E_{\lambda}({\bf k}) - \mu)) + 1)$ is the Fermi-Dirac 
distribution function with $\beta = 1/k_B T$ and $\mu$ being the chemical potential. 
We calculate the static spin susceptibility at zero temperature for hole gas by separating 
out the intraband and interband contributions as 
$\chi_{h}(0) = \chi_{\rm intra}(E_{F}) + \chi_{\rm inter}(E_{F})$. After some straight 
forward steps from Eq. (\ref{sss}), we get
\begin{eqnarray} \label{sss-intra}
\chi_{\rm intra}(E_{F}) = \frac{1}{9V} \sum_{\lambda,{\bf k} } 
|\la\lambda_1,{\bf k}|J_{z}|\lambda_2,{\bf k}\ra|^2 
\delta(E_{\lambda}({\bf k}) - E_{F})
\end{eqnarray}
and
\begin{eqnarray} \label{sss-inter}
\chi_{\rm inter}(E_{F}) = \frac{1}{9V}\sum_{\lambda_1\neq\lambda_2,{\bf k}} 
|\la\lambda_1,{\bf k}|J_{z}|\lambda_2,{\bf k}\ra|^2 
\frac{\Theta(E_F - E_{\lambda_1}({\bf k} )) -
	\Theta(E_F - E_{\lambda_2}({\bf k}))}{E_{\lambda_2}({\bf k}) - E_{\lambda_1}({\bf k})}.
\end{eqnarray}
The intraband and interband matrix elements 
$ \la\lambda_{1},{\bf k}|J_{z}|\lambda_{2},{\bf k}\ra $ in terms of polar angle 
$\theta$ of ${\bf k}$
are as folllows:
$ \la \lambda,{\bf k}|J_{z}| \lambda,{\bf k}\ra = \lambda\cos\theta,
\la \pm 1/2,{\bf k}|J_{z}| \mp 1/2,{\bf k}\ra  = -\sin\theta,
\la \pm 3/2,{\bf k}|J_{z}| \pm 1/2,{\bf k}\ra  = -(\sqrt{3}/2) \sin\theta,
\la \pm 1/2,{\bf k}|J_{z}| \pm 3/2,{\bf k}\ra  = -(\sqrt{3}/2) \sin\theta,
\la \pm 3/2,{\bf k}|J_{z}| \mp 3/2,{\bf k}\ra = 0,
\la \pm 1/2,{\bf k}|J_{z}| \mp 3/2,{\bf k}\ra = 0 $, and
$ \la \pm 3/2,{\bf k}|J_{z}| \mp 1/2,{\bf k}\ra = 0 $.

We calculate the static spin susceptibility using the above matrix 
elements in Eqs. (\ref{sss-intra}) and (\ref{sss-inter}). 
For intraband contribution to static spin susceptibility 
$\chi_{\rm intra}(E_{F}) = \chi_{\frac{1}{2}}(E_{F}) + \chi_{\frac{3}{2}}(E_{F})$, 
where contribution from light hole band $\chi_{1/2}(E_{F})$ and heavy hole band 
$\chi_{3/2}(E_{F})$ are obtained as  
$\chi_{\frac{1}{2}}(E_{F}) = \frac{3 m_{l}k_{F}^{l}}{36\pi^2\hbar^2} $ and
$\chi_{\frac{3}{2}}(E_{F}) = \frac{3m_{h}k_{F}^{h}}{36\pi^2\hbar^2} $.
For interband contribution, 
$\chi_{\rm inter}(E_{F}) = \chi_{\frac{1}{2},\frac{3}{2}}(E_{F}) 
+ \chi_{\frac{3}{2},\frac{1}{2}}(E_{F})$, 
where 
\be
\chi_{\frac{1}{2},\frac{3}{2}}(E_{F}) = \chi_{\frac{3}{2},\frac{1}{2}}(E_{F})
= \frac{1}{9\pi^2\hbar^2}\frac{(k_{F}^{h}-k_{F}^{l})m_{h}m_{l}}{m_{h}-m_{l}}.
\ee
Using the expressions of the heavy and light hole Fermi wave vectors 
$k_{F}^{h/l}$ given in Eq. (\ref{FerVec}), 
the final form of total static spin susceptibility is\cite{dietl0}
\begin{align} \label{sss-h}
\chi_{h}(0 ) & = \frac{1}{4}\rho(E_{F})
\Big[\frac{1}{3}  + \frac{8}{9}\frac{(m_{h}^{3/2} m_l - m_{l}^{3/2}m_h)}{(m_{h}-m_{l})(m_{h}^{3/2}+m_{l}^{3/2})}\Big].
\end{align}
Here $\rho(E_{F}) = (m_{h}^{3/2}+m_{l}^{3/2})^{2/3}k_{F}^{0e}/(\pi^2\hbar^2)$ is the 
density of states at the Fermi energy. 


\section{Limiting case}
The Luttinger Hamiltonian reduces to the Hamiltonian of a 3DHG with four
degenerate bands by setting $\gamma_1=1 $ and $\gamma_s=0$. 
In this limit, the static hole spin susceptibility calculated from Eq. (\ref{sss-h}) 
is
\begin{align} \label{chi-limit}
\tilde \chi_{h} = \frac{10m_0 k_{F}^{0h}}{36\pi^2\hbar^2}.
\end{align}
Similarly, setting $\gamma_1 = 1 $ and $\gamma_s = 0$ in Eqs. \ref{J_H} and \ref{J_I}, 
the Ising-like range function exactly vanishes i.e . $J_I=0$ and
the Heisenberg-like range function $J_H$ becomes
\begin{align} \label{Jh-limit}
J_H & = - \frac{2J_{pd}^{2}(k_{F}^{0h})^3}{\pi} \tilde \chi_h 
\mathcal{F}(2k_{F}^{0h}R),
\end{align}
where $\mathcal{F}(y) = \frac{\sin(y) - y \cos(y)}{y^4}$.
On the other hand, the analytical form of the RKKY interaction for
conventional 3DEG with two degenerate bands is
given by \cite{diet}
\begin{align} \label{Jh-limit1}
J_H & = - \frac{2J_{sd}^{2}(k_{F}^{0e})^3}{\pi} \tilde \chi_e 
\mathcal{F}(2k_{F}^{0e}R),
\end{align}
where $\tilde \chi_{e} = m_0 k_{F}^{0e}/(4\pi^2\hbar^2)$ is the static 
electron spin susceptibility.
So we see from Eqs. (\ref{Jh-limit}) and (\ref{Jh-limit1}) that similar relation between RKKY 
interaction and static spin susceptibility follows for both the cases. 


\end{widetext}

\end{document}